\documentclass[a4paper,11pt]{scrartcl}
\usepackage{ILD}
\usepackage{tagging}
\usetag{ILD}
\usepackage{subcaption}
\usepackage{hyperref}
\usepackage{wasysym}
\usepackage{slashed}
\usepackage{wrapfig,rotating}
\usepackage{pgfpages}
\usepackage{fancybox,graphicx}
\usepackage{graphbox}
\usepackage{epstopdf}
\AppendGraphicsExtensions{.eps.gz}
\title{
   New physics searches with the International Large Detector at the ILC

}
\ildproc{phys}{2021}{007}
\date{\today}

\addauthor{
  Mikael Berggren

}
{\institute{1}}

\addinstitute{1}{
    Deutsches Elektronen-Synchrotron DESY,\\
  Notkestr. 85, 22607 Hamburg, Germany

}
\onbehalfof{
    \tagged{POS}{On behalf of }the ILD concept group

}

\abstract{
  Although the LHC experiments have searched for and excluded many proposed new particles
up to masses close to 1 TeV, there are many scenarios that are difficult to address at a
hadron collider.  This talk will review a number of these scenarios and present the
expectations for searches at an electron-positron collider such as the International Linear
Collider.   The cases discussed include the light Higgsino, the \stau~ slepton in the coannihilation
region relevant to dark matter, as well as other SUSY signatures.
The studies are based on the ILD concept at the ILC.

}

\titlecomment{%
    *** The European Physical Society Conference on High Energy Physics (EPS-HEP2021), ***\\
  *** 26-30 July 2021 ***\\
  *** Online conference, jointly organized by Universität Hamburg and the research center DESY ***

}

\def\leqsim{\mathbin{\;\raise1pt\hbox{$<$}\kern-8pt\lower3pt\hbox{$\sim$}\;}}
\def\geqsim{\mathbin{\;\raise1pt\hbox{$>$}\kern-8pt\lower3pt\hbox{$\sim$}\;}}

\def\MXN#1{\mbox{$ M_{\tilde{\chi}^0_#1}                                $}}
\def\MXC#1{\mbox{$ M_{\tilde{\chi}^{\pm}_#1}                            $}}

\def\XPM#1{\mbox{$ \tilde{\chi}^{\pm}_#1                                $}}

\def\XN#1{\mbox{$ \tilde{\chi}^0_#1                                     $}}

\def\p#1{\mbox{$ \mbox{\bf p}_1                                         $}}

\newcommand{\smu}     {\mbox{$ \tilde{\mu}                                 $}}

\newcommand{\sel}     {\mbox{$ \tilde{\mathrm e}                           $}}

\newcommand{\stau}    {\mbox{$ \tilde{\tau}                                $}}
\newcommand{\stone}   {\mbox{$ \tilde{\tau}_1                              $}}

\newcommand{\GeV}     {\mbox{$ {\mathrm{GeV}}                              $}}

\newcommand{\TeV}     {\mbox{$ {\mathrm{TeV}}                              $}}

\newcommand{\ba}{\begin{array}}
\newcommand{\ea}{\end{array}}
\newcommand{\bc}{\begin{center}}
\newcommand{\ec}{\end{center}}
\newcommand{\be}{\begin{eqnarray}}
\newcommand{\eeq}{\end{eqnarray}}
\newcommand{\bes}{\begin{eqnarray*}}
\newcommand{\ees}{\end{eqnarray*}}
\newcommand{\Kz}{\ifmmode {\rm K^0_s} \else ${\rm K^0_s} $ \fi}
\newcommand{\Zz}{\ifmmode {\rm Z^0} \else ${\rm Z^0 } $ \fi}
\newcommand{\xxbar}{\ifmmode {\rm x\bar{x}} \else ${\rm x\bar{x}} $ \fi}
\newcommand{\rphi}{\ifmmode {\rm R\phi} \else ${\rm R\phi} $ \fi}

\def    \missEt      {\ifmmode{/\mkern-11mu E_t}\else{${/\mkern-11mu E_t}$}\fi}
\def    \missE       {\ifmmode{/\mkern-11mu E}\else{${/\mkern-11mu E}$}\fi}
\def    \missp       {\ifmmode{/\mkern-11mu p}\else{${/\mkern-11mu p}$}\fi}
\def    \misspt      {\ifmmode{/\mkern-11mu p_t}\else{${/\mkern-11mu p_t}$}\fi}

\begin{document}
\titlepage

\section{Introduction: ILC and ILD and their strong points for searches}

\begin{figure}[h] 
  \begin{center}
  \includegraphics [scale=0.3]{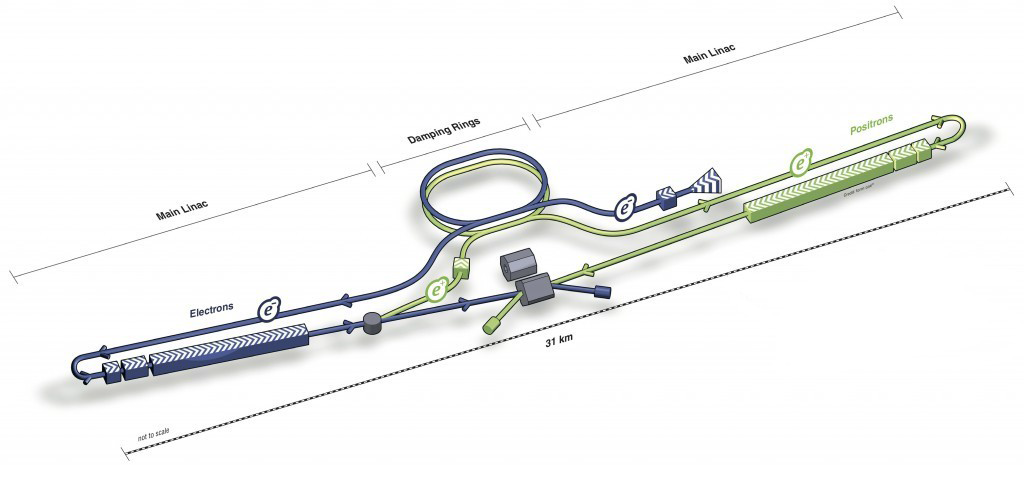}
  \includegraphics [scale=0.3]{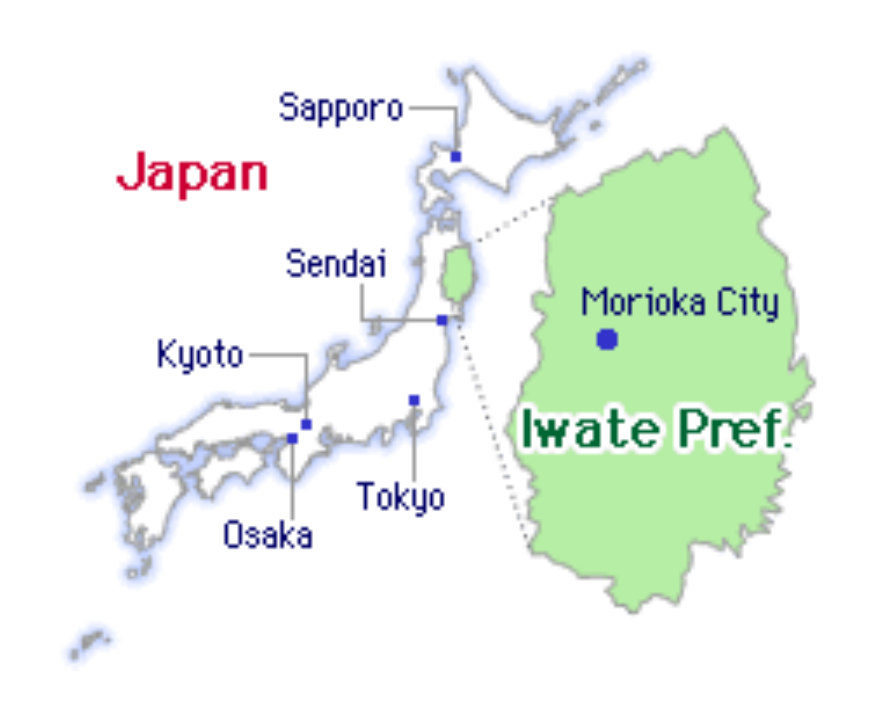}
   \end{center}
 \caption{Schematic of the ILC and the location of the proposed site in Japan's Tohoko region.\label{fig:ilc}}
 \end{figure}

The International Linear Collider (the ILC, Fig.~\ref{fig:ilc}) will collide polarised electrons with polarised positrons.
Centre-of-mass energies will range from 250 \GeV~ to 500 \GeV, with an upgrade path to 1 \TeV~
defined.
An $E_{CMS}= M_Z$ option is also foreseeable.
As the $e^+e^-$ initial state implies electroweak production, the background rates will be quite low.
This has consequences for the detector design and optimisation: The detectors can feature close to $ 4\pi$ coverage,
and they do not need to be  radiation hard, so that the tracking system in front of calorimeters
can have a thickness as low as a few percent of a radiation-length.
In addition, the low rates means that the detectors needn't be triggered, so that \textit{all}
produced events will be available to analysis.
Furthermore, at an   $e^+e^-$ machine, point-like objects are brought into  collision,
meaning that the initial state is fully  known.

The  ILC \cite{Adolphsen:2013kya} has a defined 20 year running scenario, yielding
integrated luminosities of 2 and 4 ab$^{-1}$  at $E_{CMS}=$ 250 and 500 \GeV, respectively,
and could deliver 8  ab$^{-1}$ at the possible upgrade to 1 \TeV.  
To construct the ILC is currently under high-level political consideration in Japan.

 The excellent conditions provided by the accelerator need to be matched by excellent
 performance of the detectors.
 Both precision SM measurements as well as Beyond the Standard Model (BSM) searches or measurements will require
 a jet energy resolution of  3-4\%,
   \begin{wrapfigure}{r}{0.5\columnwidth}
    \begin{center}
    \includegraphics [scale=0.085]{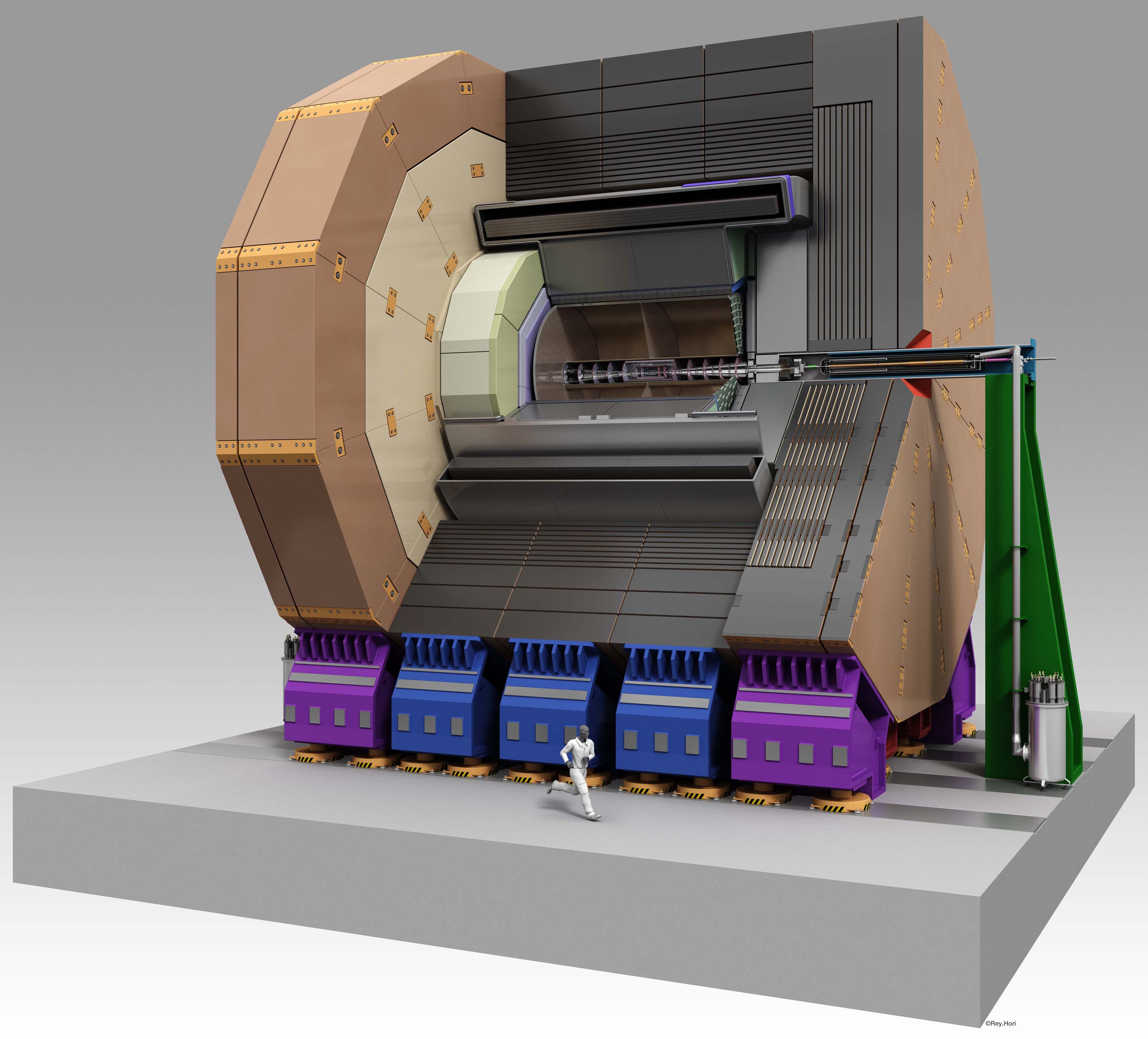}
    \end{center}
 \caption{Artist's view of the ILD concept\label{fig:ild}.}
\end{wrapfigure}     
 an asymptotic  momentum resolution of $\sigma(1/p_\perp) = 2 \times 10^{-5}$ \GeV $^{-1}$,
and measurement of impact-parameters better than 5 $\mu$ .
 In addition, powerful particle identification capabilities are needed.
 The detector should be hermetic, with the only gaps in acceptance being the
 unavoidable vacuum pipes bringing the beams into the detector.
 Furthermore, the system should be capable to register data without any trigger.

 In the ILD concept \cite{ILDConceptGroup:2020sfq}, illustrated in Fig.~\ref{fig:ild}, a low mass,
 high precision, tracker
 with PID capabilities  is achieved by having a Time Projection Chamber
 as the main tracker, enhanced by silicon trackers both inside and
 outside the TPC.
 To achieve the needed jet energy resolution,
 ILD incorporates
 high granularity calorimeters optimised for particle flow.
 The entire system can be operated in 
 power-pulsing mode, i.e.~with the electronics being switched off
 between bunch-trains. In this mode, no active cooling will be needed.

    
  \section{BSM at ILC: The SUSY case}

  In this short contribution, we concentrate
  on one theory for physics beyond the standard model, namely
  SUSY\cite{Wess:1974tw}.
   Not only is SUSY the most complete theory of BSM, it can also
   serves as a boiler-plate for BSM in general, since almost any new topology can
   be obtained in some flavour of SUSY, in particular if also possible violation of R-parity and/or CP-symmetry,
   or non-minimal models
   are considered.
   In addition, it is the paradigm  that has been most studied
   with detailed detector simulation. In most cases, studies were done with
   full simulation of the ILD, with all SM backgrounds, and all
      beam-induced backgrounds included.
    It is true that SUSY is under some stress by recent
    LHC results. However, ILC offers different angles to explore the properties
    of SUSY compared to LHC,
    e.g. loop-hole free searches, and
    complete coverage of compressed spectra.

    Naturalness, the hierarchy problem, the nature of dark matter (DM),
  or the observed value of the magnetic moment of the muon, all prefer
  a light electroweak sector of SUSY.
  Except for the third generation squarks, the coloured sector
  - where pp machines excel -
  does not provide any insight into any of these issues.
  In addition, many models point into this direction:
  If the Lightest SUSY Particle (the LSP) is Higgsino or Wino, there must be other
  bosinos close in mass to the LSP, since the $\tilde{H}$ and $\tilde{W}$
  fields have several components, leading to a close relation between
  the physical bosino states;
  only a Bino-LSP can have large difference, $\Delta(M)$, between the LSP and the
  Next to Lightest
  SUSY Particle (the NLSP).
  Furthermore,
  if the LSP is Higgsino, one can obtain \textit{Natural SUSY}:
  In such models one finds
  that requiring low fine-tuning leads to
  the condition that the Higgsino mass-parameter 
  $\mu$ must be  $\mathcal{O}(m_Z)$, i.e. an LSP at the weak scale.
  In the case of such compressed, low $\Delta(M)$, spectra, most sparticle-decays are
  via cascades,
  where the last decay in the cascade - that to SM particles and the LSP -
  features small $\Delta(M)$.
  For such decays, current LHC limits are for specific models,
  and only the limits from LEPII are model-independent.
In fact, current observations  from LHC13, LEP, g-2, DM (assumed to be 100\%~LSP),
and precision observables
taken together also point to a compressed spectrum \cite{Bagnaschi:2017tru}.
\begin{figure}[t]
   \begin{center}
     \includegraphics [align=t,scale=0.30]{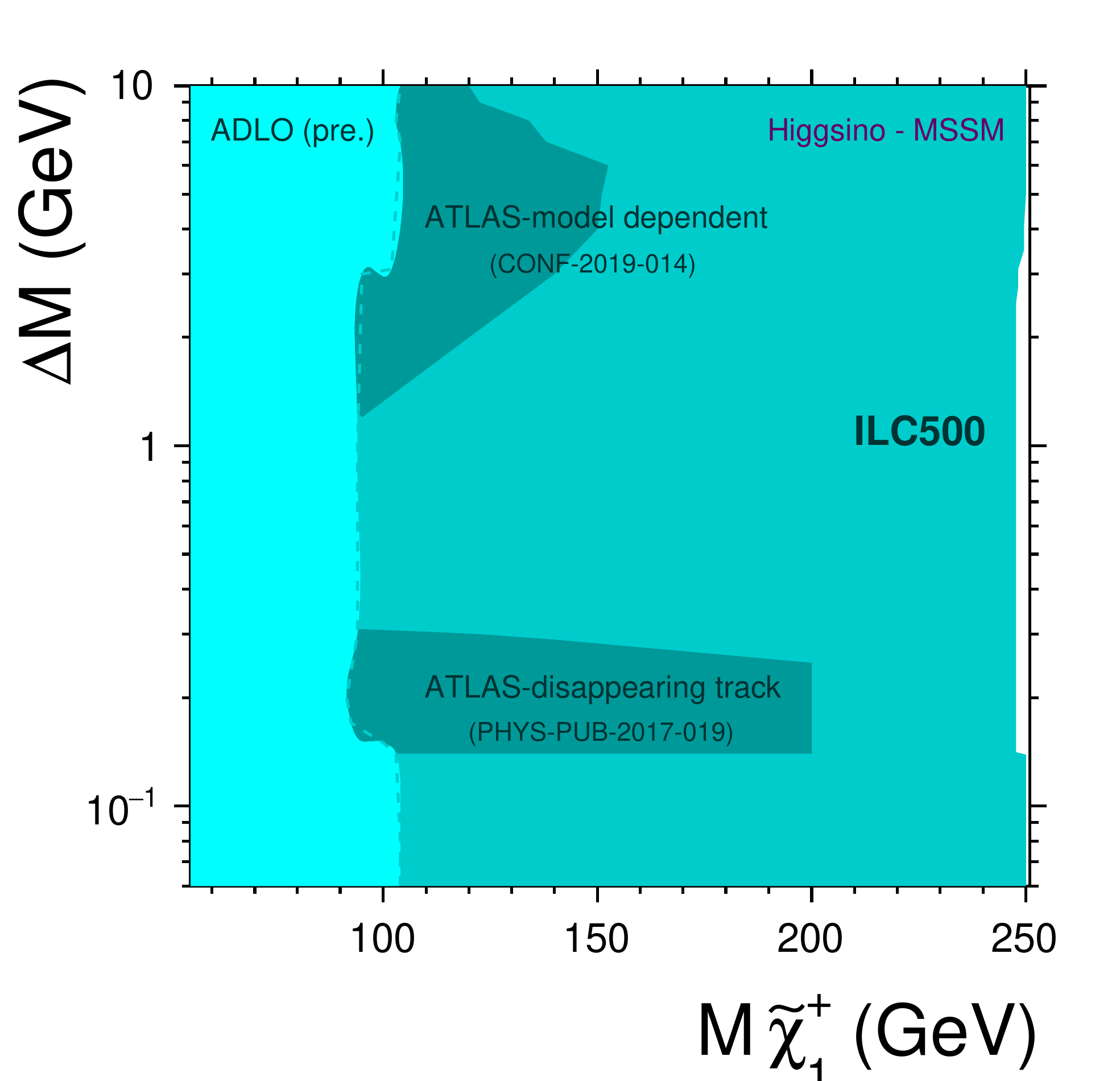}
     \includegraphics [align=t,scale=0.27]{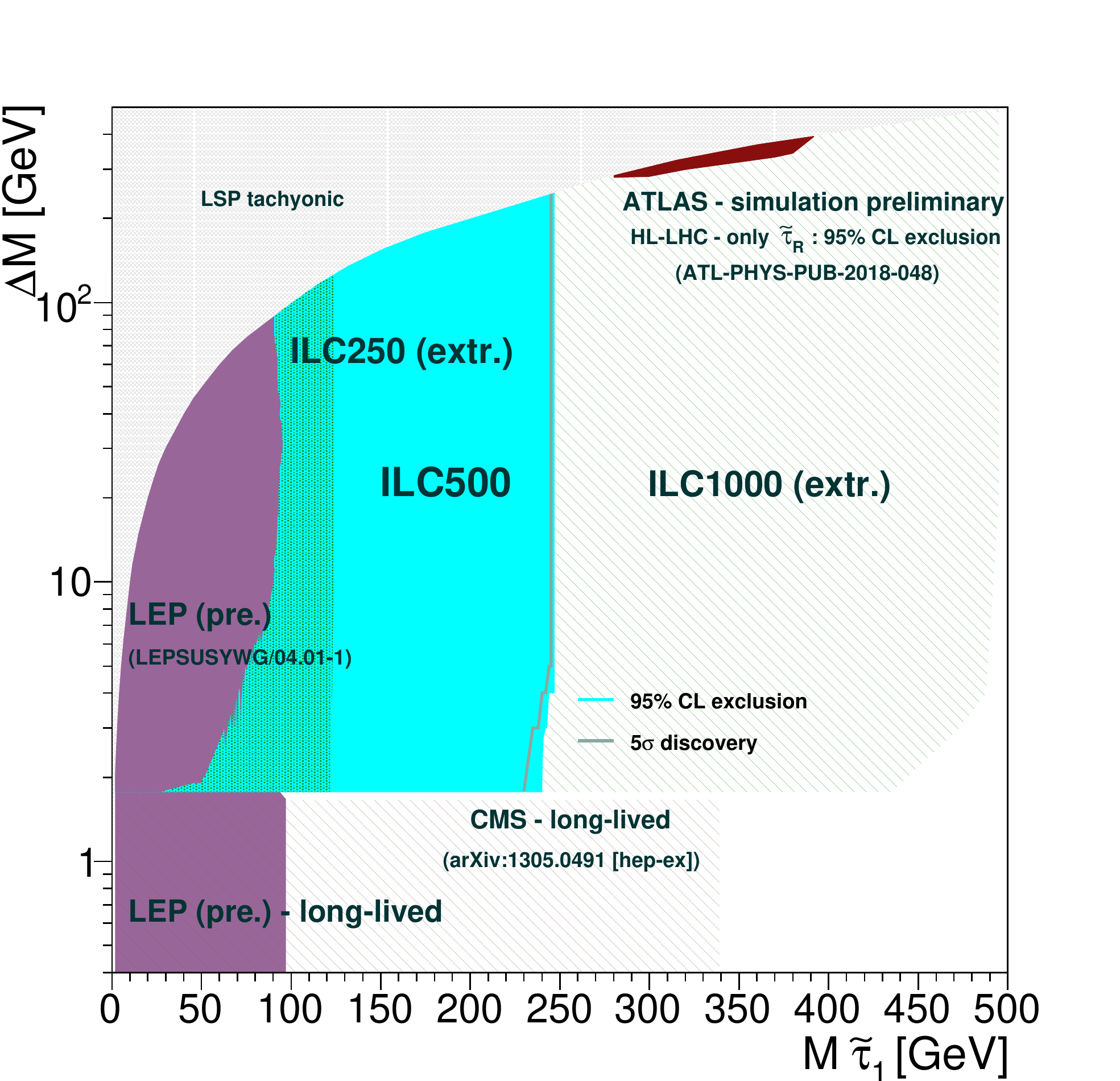}
\end{center}
   \caption{ Exclusion and discovery reaches for a $\XPM{1}$ (left), or a $\stone$ (right).\label{fig;C1stauexcl}}
\end{figure} 
  \begin{figure}[b]
    \begin{center}
\includegraphics [scale=0.38]{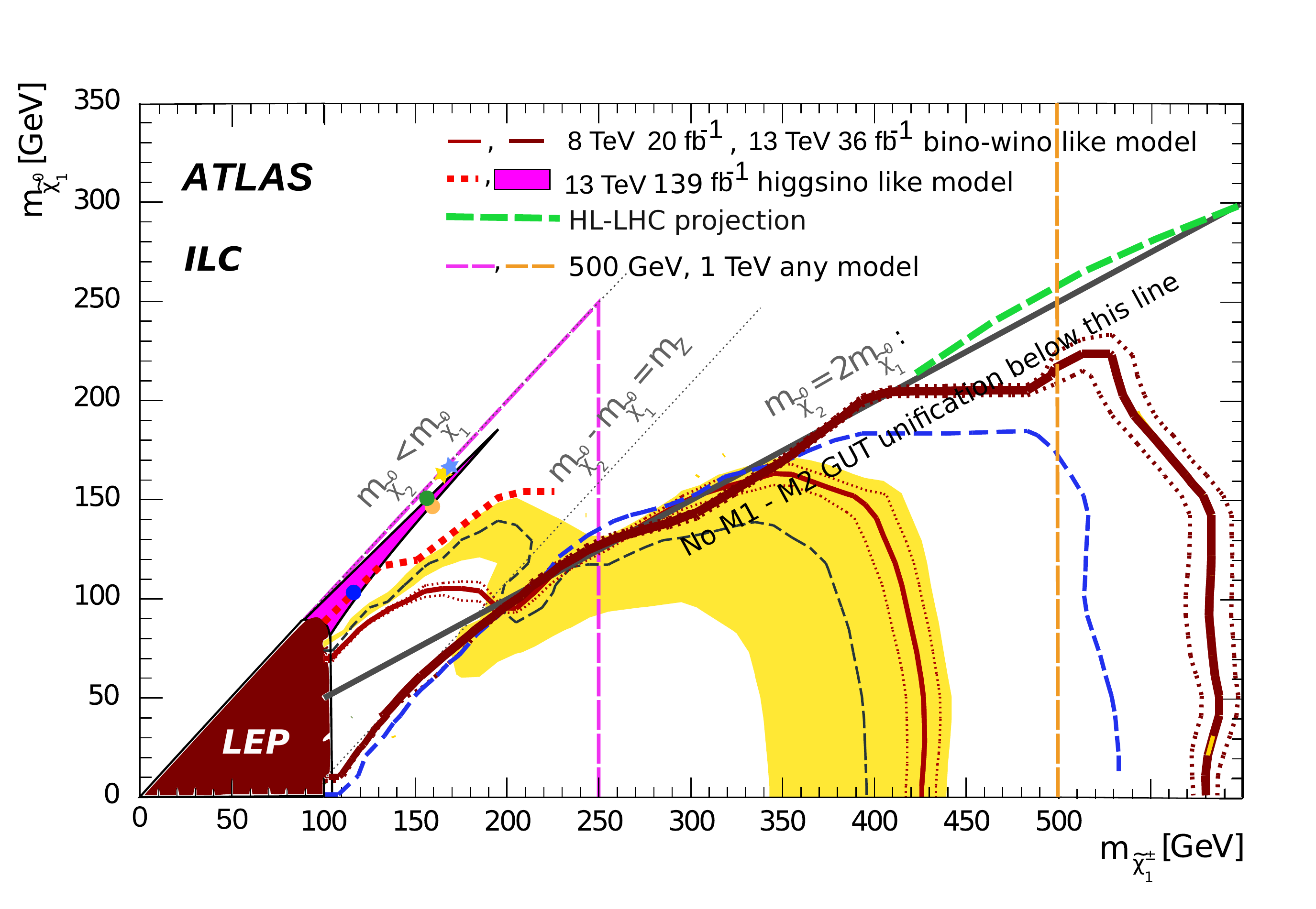}
\end{center}
\caption{ Observed or projected exclusion regions for $\XPM{1}$ NLSP, for LEPII, LHC, HL-LHC and for ILC-500 and ILD-1000
  The symbols indicate where the higgsino LSP models shown in Fig. \ref{fig:sleptC1N2} are located.\label{fig:X1summary}}
 \end{figure}

\subsection{SUSY with no loop-holes}
At ILC one can perform a loophole free search for SUSY
because in SUSY, the properties of NLSP
production and decay are completely predicted
  for given LSP and NLSP masses, due to
  SUSY-principle: ``sparticles couples as particles''.
Note that this does not depend on the (model dependent) SUSY breaking mechanism.
  By definition, there is only one NLSP, and it must have 100\% BR
  to its (on- or off-shell) SM-partner and the (stable or unstable) LSP.
  Also, there is only a handful of possible candidates to be the NLSP.
  Hence by performing searches for every possible NLSP, 
  model independent exclusion and  discovery reaches in the $M_{NLSP} - M_{LSP}$ plane,
separately for each NLSP candidate, or globally, by determining which NLSP gives the
weakest limit at any point. There will be no loopholes to the conclusion.
Examples of this procedure are shown in Fig. \ref{fig;C1stauexcl} for the cases of a $\XPM{1}$ \cite{PardodeVera:2020zlr} or
a $\stone$ \cite{NunezPardodeVera:2021cdw} NLSP.
The $\XPM{1}$ is a conservative extrapolation from the LEP results, while the
$\stone$ one is obtained with detailed fast simulation of ILD, where the $\stau$
and LSP properties were chosen such that the limit is the weakest possible one,
i.e. the experimentally ``worst possible'' case.
In the figure, it can be seen that the discovery and exclusion reaches are almost
the same, and reach quite close to the kinematic limit $2 M_{NLSP} = E_{CMS}$.
It should also be noted that the HL-LHC projection from ATLAS is exclusion only,
and is for specific assumptions on the $\stau$ properties, assumptions that are
not the most pessimistic. In Fig. \ref{fig:X1summary}, the various current or projected limits are
shown in a single plot \cite{PardodeVera:2020zlr,ATLAS:2018ojr}.
It should be noted that below the heavy black line, GUT unification of the
Bino and Wino mass-parameters  $M_1$ and $M_2$
is not possible: The difference between $\MXN{1}$ and $\MXC{1}$ cannot be larger than what the line indicates,
if such a unification is realised.
\tagged{POS}{
  
 {\bf 2.2  ~~SUSY at ILC: discovery in a week, then precision measurements.}
 \\
} 
 \begin{wrapfigure}{l}{0.35\columnwidth}
   \begin{center}
     \includegraphics [scale=0.22]{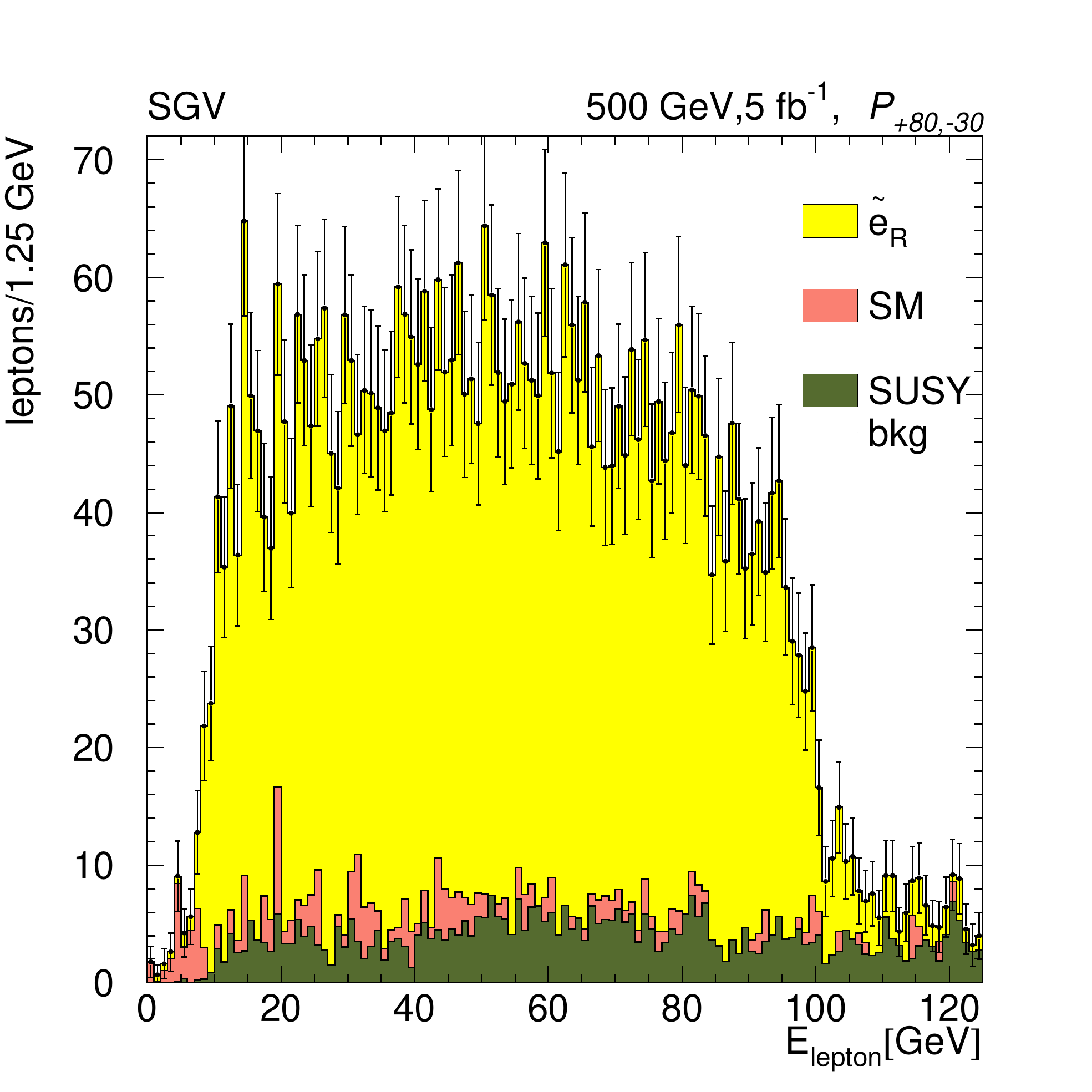}
   \end{center}
   \caption{ The $\sel$ signal after 5 fb$^{-1}$ have been collected. \label{fig:STCsel5fb}}
\end{wrapfigure}
\tagged{ILD}{
\vskip 0.1cm
\noindent
\textsf{\textbf{2.2  ~~SUSY at ILC: discovery in a week, then precision measurements.}}
\vskip 0.5cm
}  
\begin{figure}[b] 
   \begin{center}
     \includegraphics [scale=0.21]{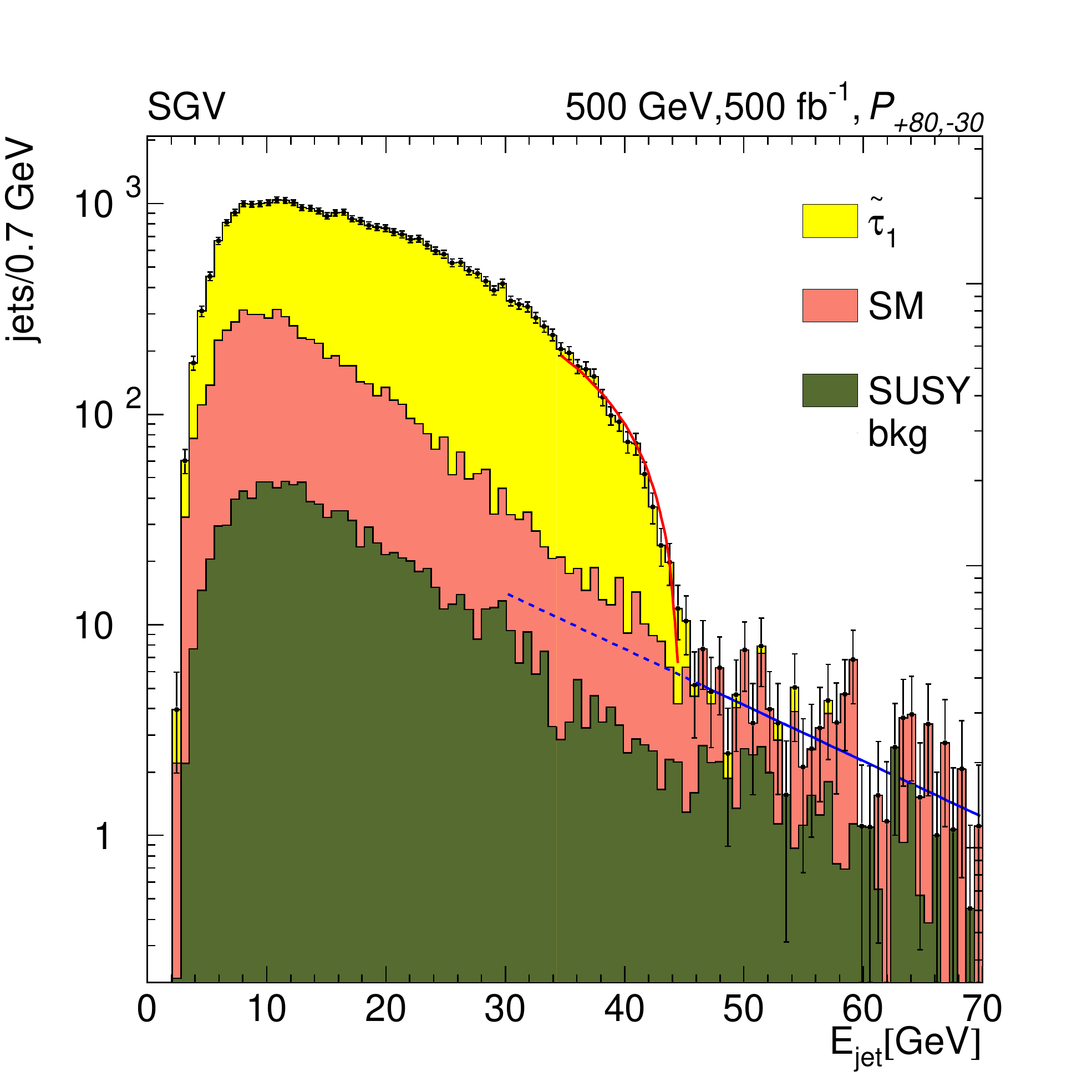}
     \includegraphics [scale=0.21]{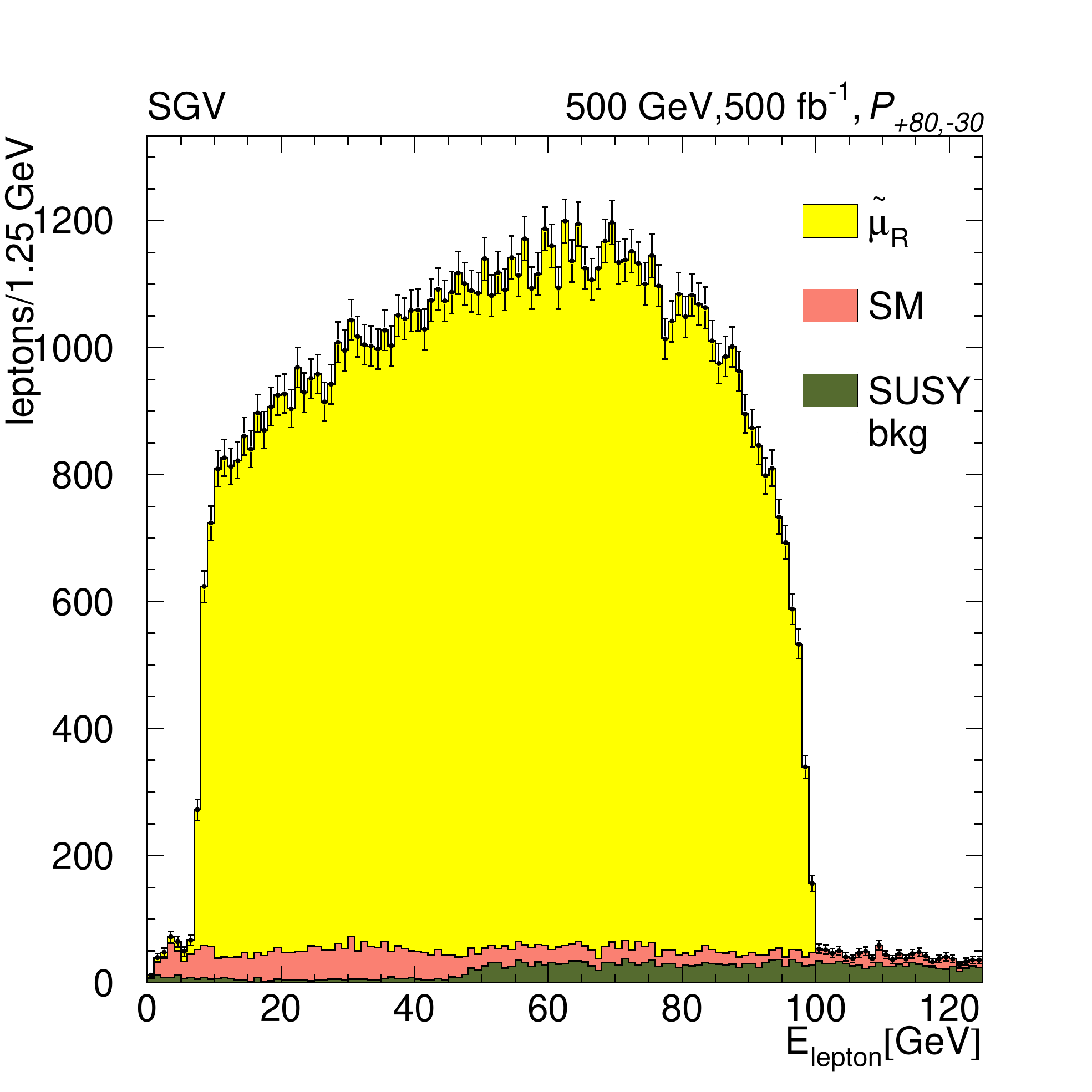}
     \includegraphics [scale=0.21]{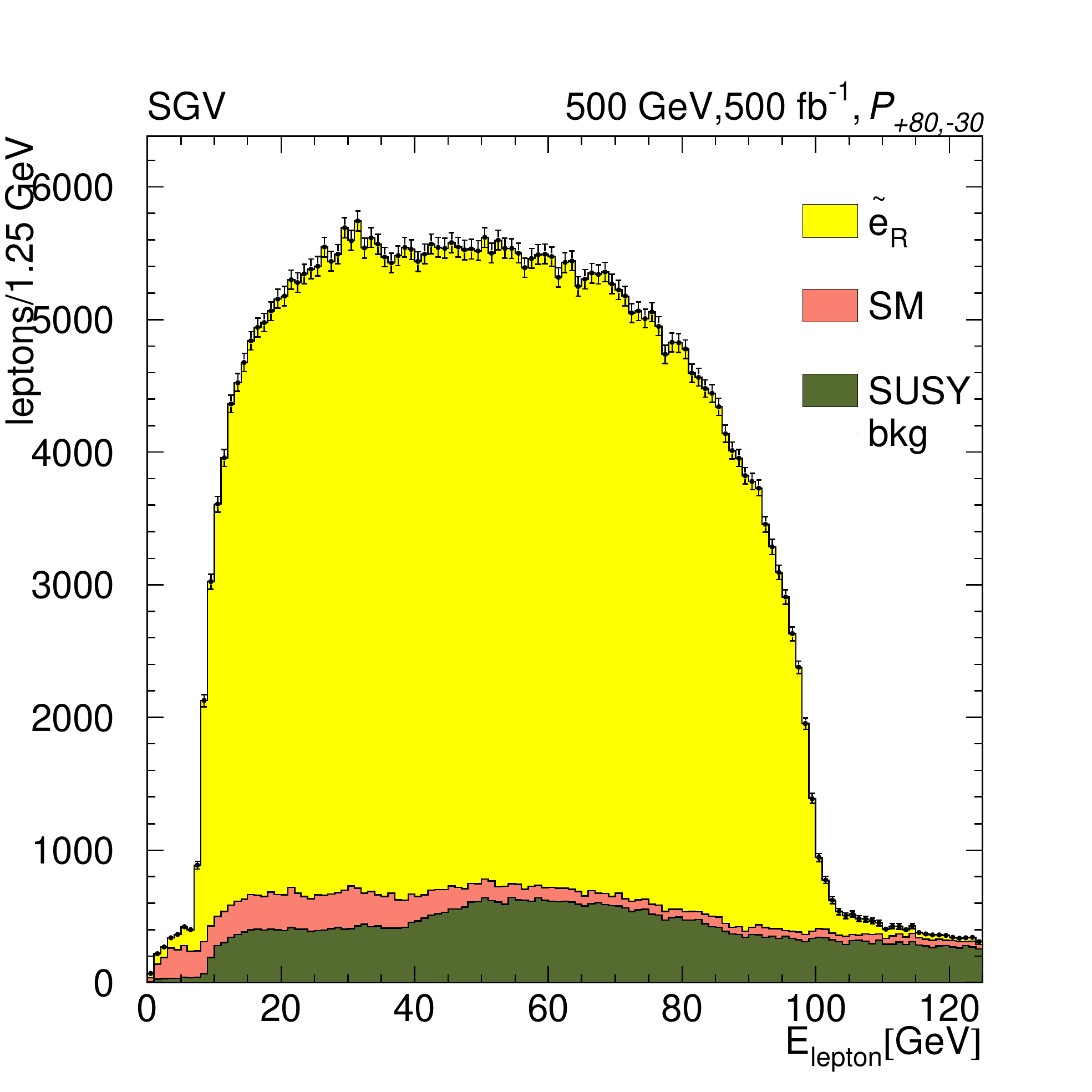}
     
      \includegraphics [align=c,scale=0.195]{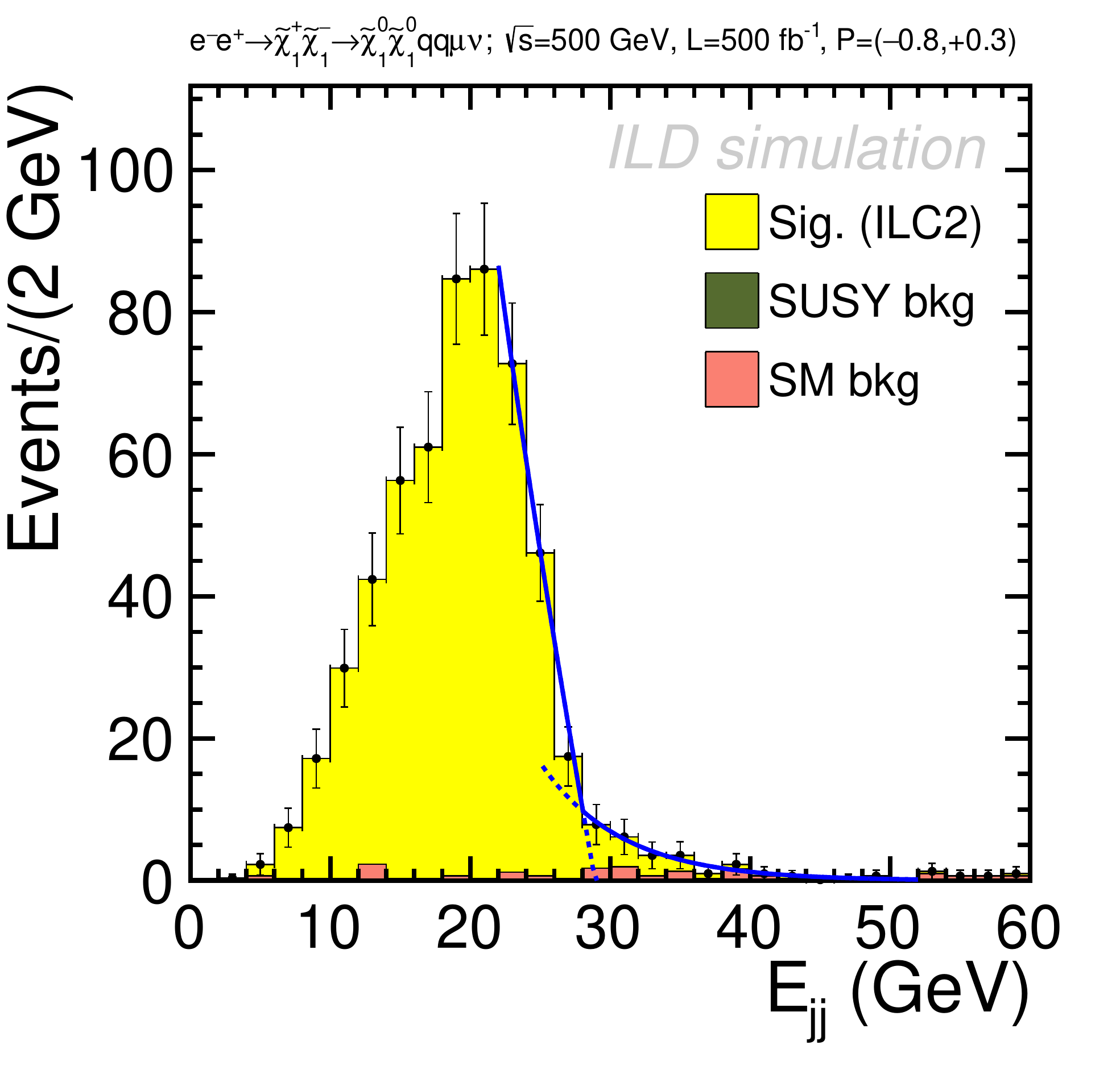}
      \includegraphics [align=c,scale=0.195]{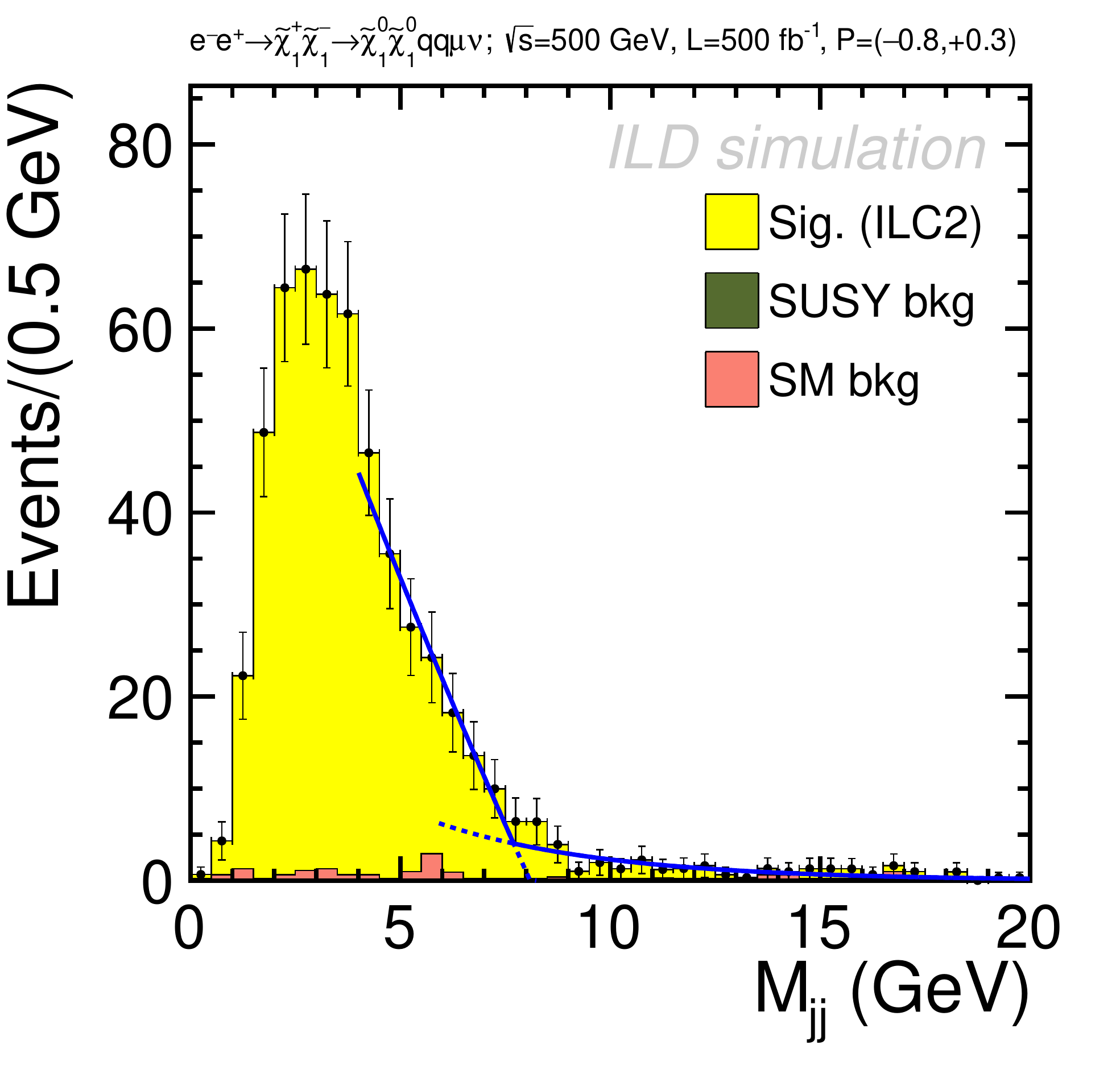}
      \includegraphics [align=c,scale=0.21]{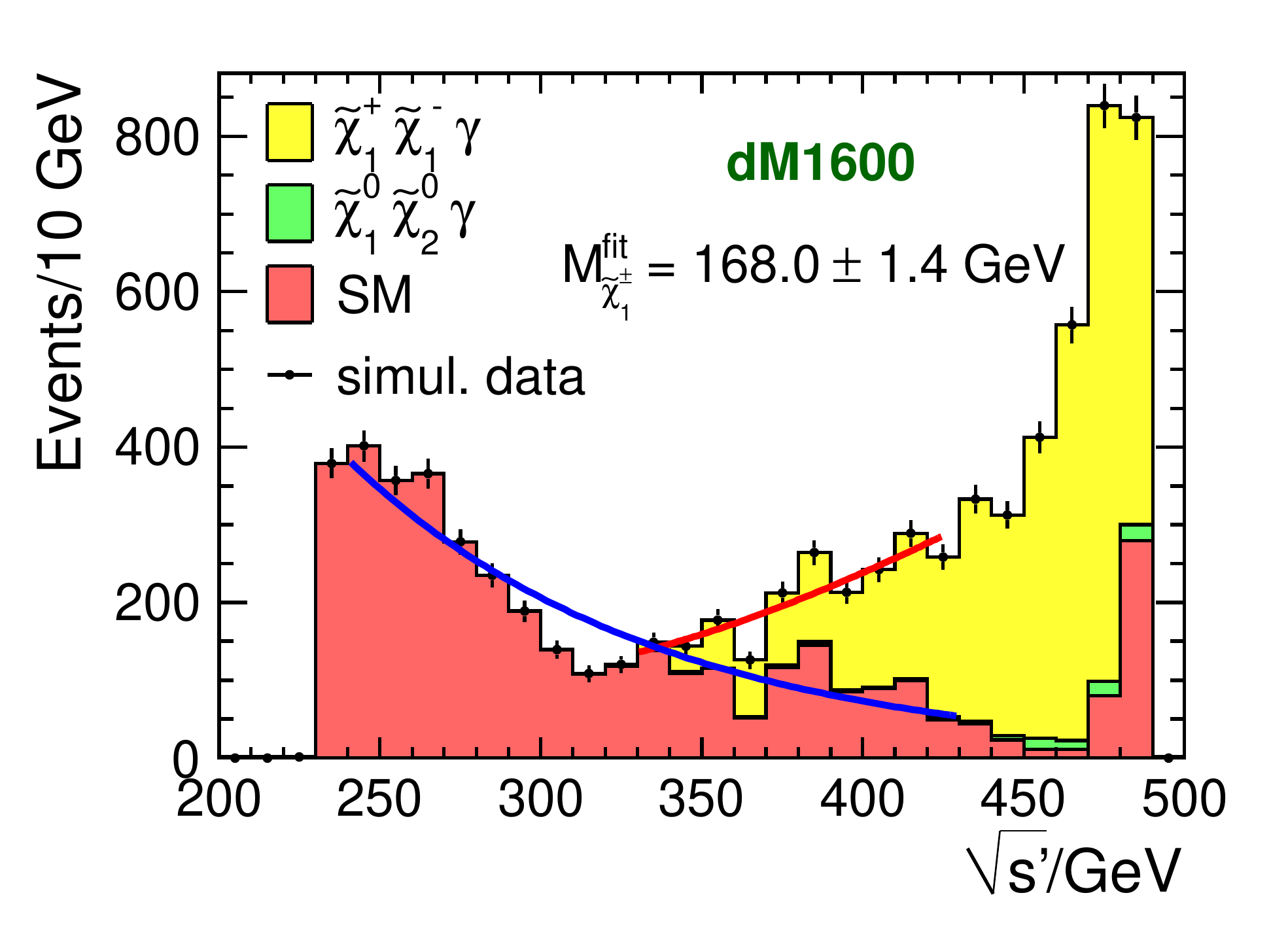}
      
      \includegraphics [align=c,scale=0.195]{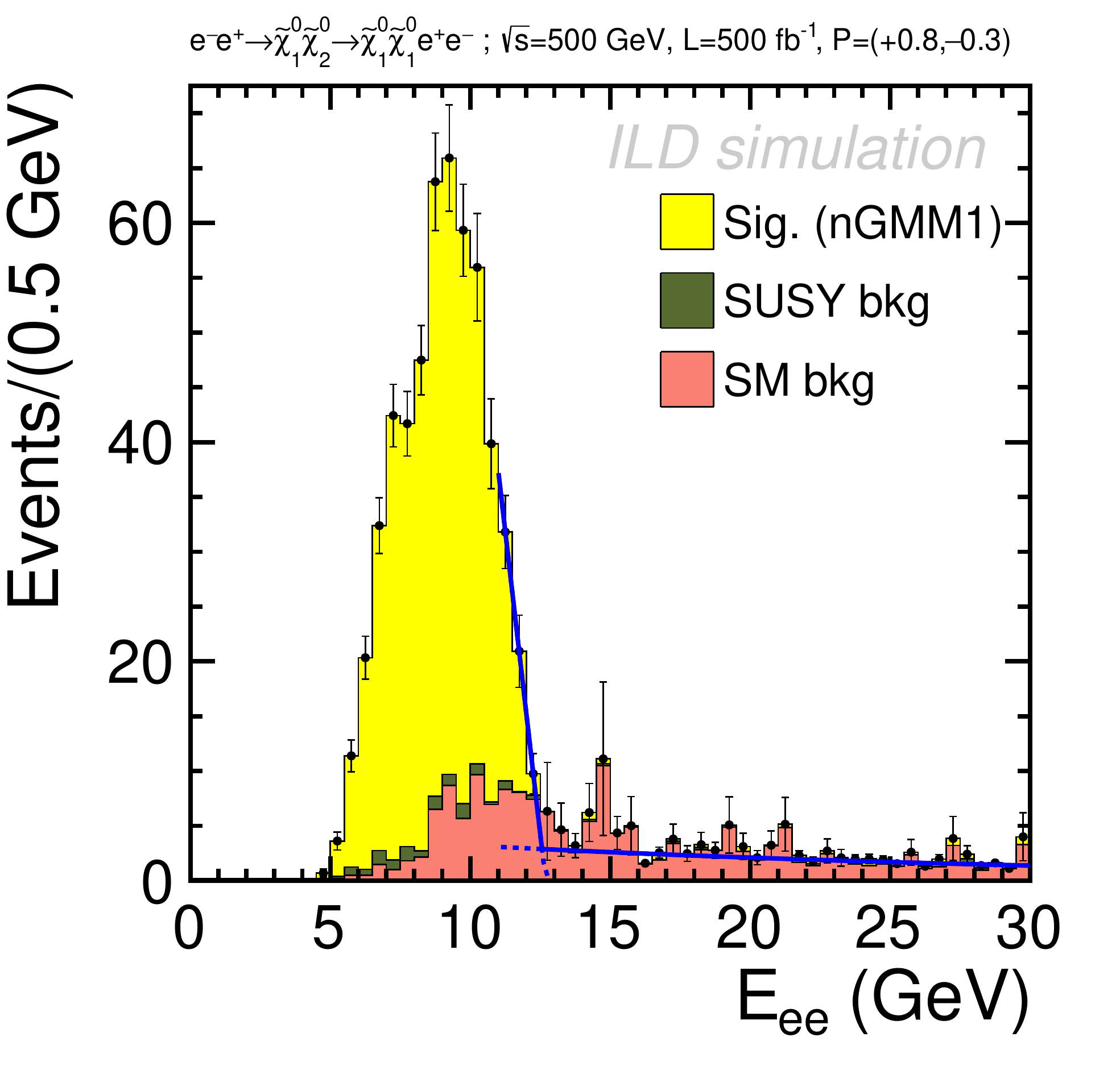}
      \includegraphics [align=c,scale=0.195]{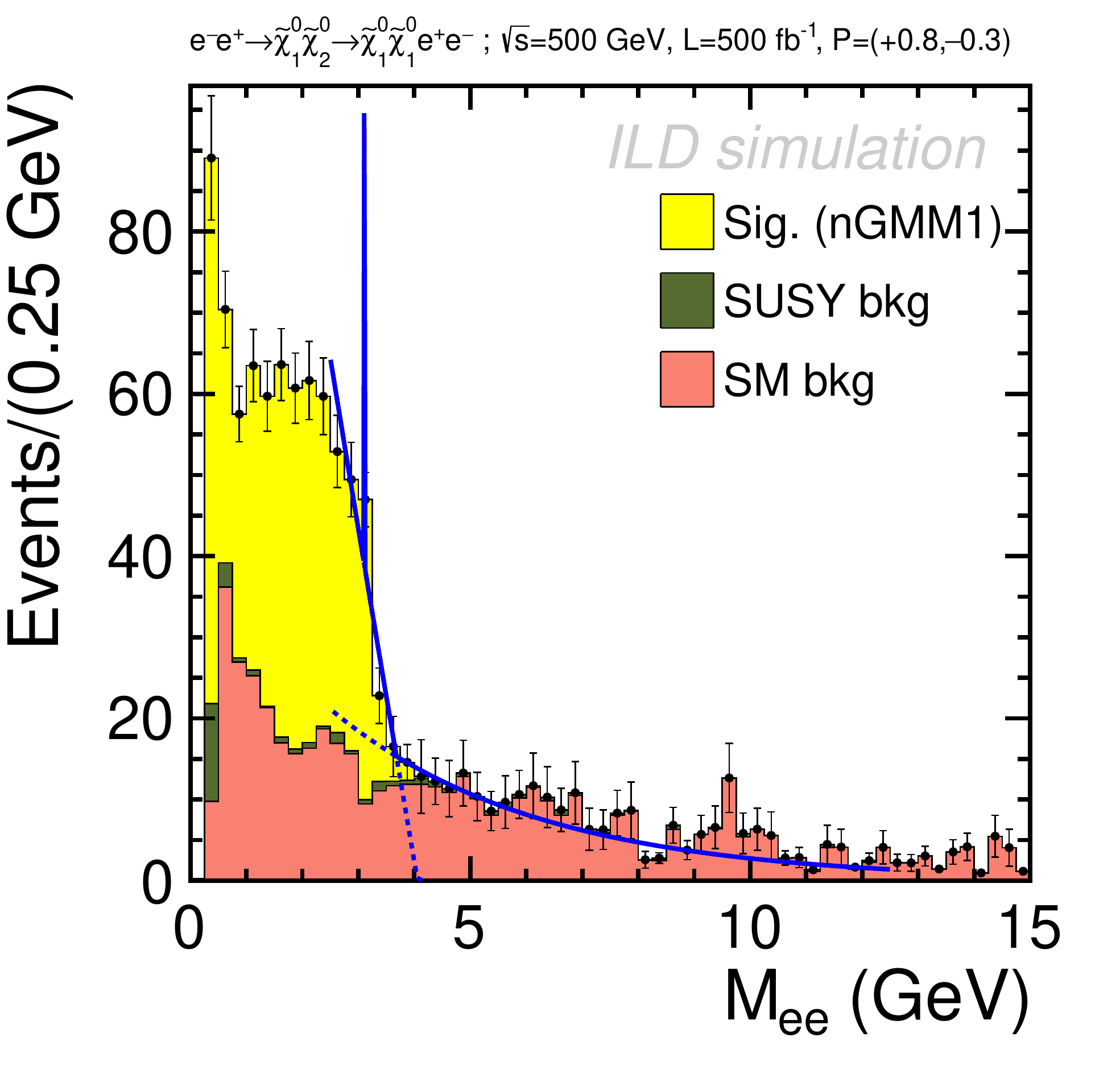}
      \includegraphics [align=c,scale=0.21]{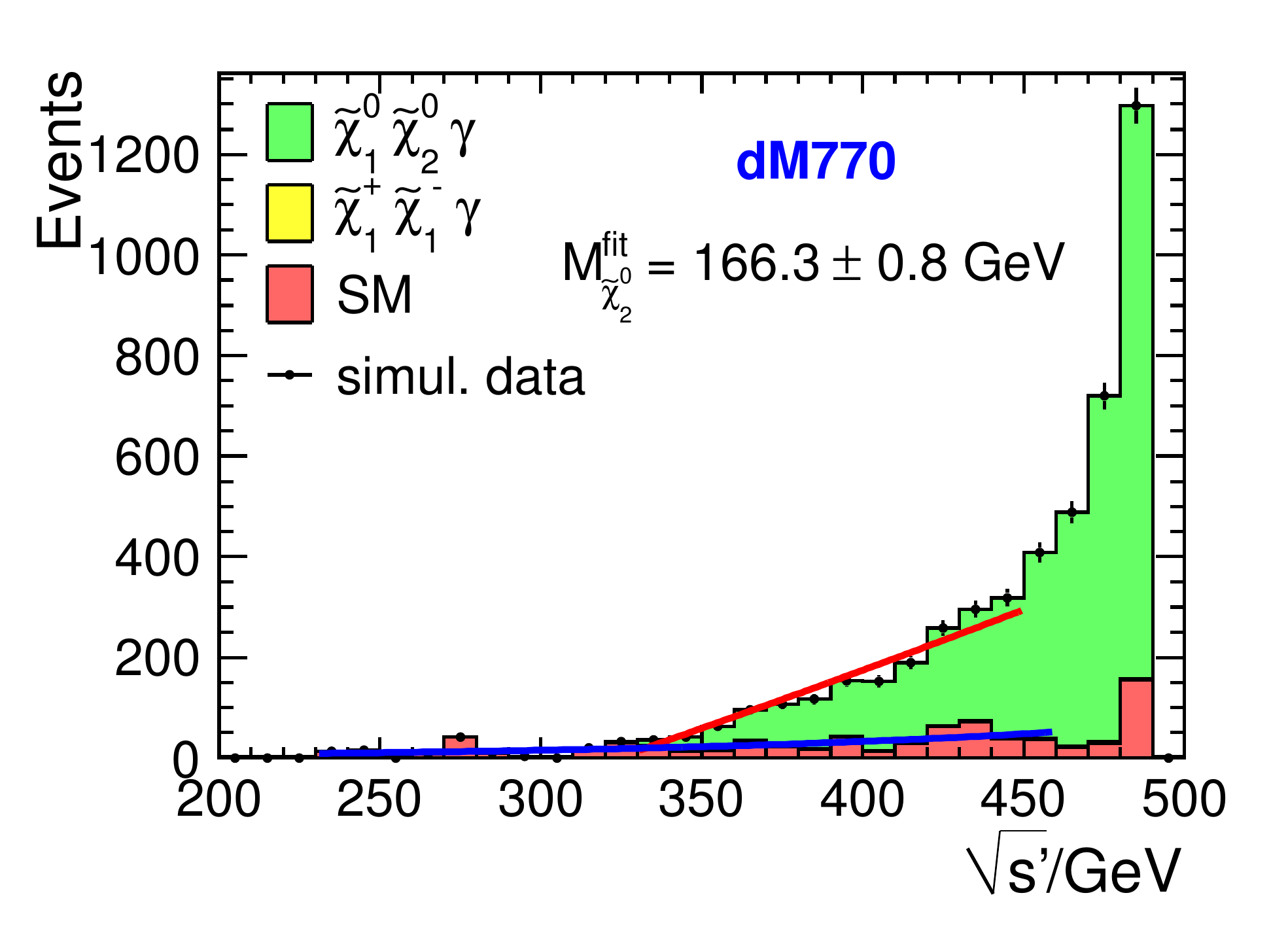}
    \end{center}
    \caption{Top row: $\stau$, $\smu$ and $\sel$ spectra. Middle and bottom rows: Observables
      for three different Higgsino-LSP models. The middle row shows the case of $\XPM{1}$ production, the bottom one 
      that of $\XN{2}$ production.  \label{fig:sleptC1N2}}
\end{figure}  
In fact, at the ILC, SUSY discovery would happen quite quickly. Fig. \ref{fig:STCsel5fb}
shows  ILD fast detector simulation studies of $\sel$ production in 
a $\stau$ co-annihilation model \cite{Berggren:2015qua}. The figure shows the signal 
after collecting only 5 fb$^{-1}$ of integrated luminosity.
Under nominal running conditions, this corresponds to  1 week of data-taking.
One sees that at ILC, the situation that a interesting SUSY signal will be at the intermediate
significance (neither excluded, nor discovered) for years will never occur: Either the
process is not in reach and there is no sign of it, or it will be discovered immediately.

This means that studies of SUSY at ILC would almost directly enter into the
realm of precision studies. The plots in Fig. \ref{fig:sleptC1N2} shows a number of examples of the
kind of signals that would be expected, obtained by
ILD detector simulation studies:
 Typical slepton signal ($\stau$, $\smu$ and $\sel$) in the top row,
       in a  $\stau$ co-annihilation model (FastSim) \cite{Berggren:2015qua}.
       Typical chargino and neutralino signals in different Higgsino LSP models are shown in the following rows.
       The left-hand two plots are models  with moderate (a few to some tens GeV) $\Delta{M}$ (FullSim) \cite{Baer:2019gvu},
       while the right-hand ones is  for a  model with very low (sub-GeV) $\Delta{M}$ (Fast/FullSim) \cite{Berggren:2013vfa}.
      In all the illustrated cases, it was found that the
 SUSY masses could be determined at the sub-percent level,
 the polarised production cross-sections to the level of a few percent.
 Many other properties could also be obtained from the same data, such as
 decay branching fractions, mixing angles, and sparticle spin.



\section{Conclusions}
    Sometimes, the capabilities for the {direct discovery} of new particles 
at the ILC exceed those
of the LHC, since ILC provides
a  well-defined initial state, and a
clean environment without QCD backgrounds.
ILC also is extendable in energy and features polarised beams. 
In addition, detectors like ILD, will be factors more precise, will be hermetic, and will not need
for to be triggered.

Many ILC - LHC synergies are expected, from the high energy-reach of LHC versus the high
sensitivity of the ILC.
In particular, for SUSY,
the high mass reach of LHC is ideally complemented by the sensitivity for  low $\Delta(M)$ at ILC. 
If SUSY is reachable at the ILC, it means 5 $\sigma$ discovery, and precision measurements.
This input
  might be just what is needed for LHC to transform a 3 $\sigma$ excess to a discovery of states
  beyond the reach of ILC.


\end{document}